\documentclass{moriond_pn}

\usepackage{amsmath}

\RequirePackage{booktabs}
\graphicspath{{figures/}}

\newcommand{\arcsec}{\hbox{$^{\prime\prime}$}}

\newcommand{\muas}{\hbox{$\mu$}as}
\newcommand{\degr}{\ensuremath{^\circ}}

\def\cent{\multicolumn{1}{c}}

\newcommand{\kms}{\,km\,s$^{-1}$}

\def\be{\begin{equation}}
\def\ee{\end{equation}}
\def\bea{\begin{eqnarray}}
\def\eea{\end{eqnarray}}

\newcommand{\Photo}{}

\begin{document}
\vspace*{4cm}
\title{The Gaia MISSION and SIGNIFICANCE\ \footnote{To appear in Proceedings of Rencontres de Moriond 2019, Gravitation}}

\author{F. MIGNARD }

\address{Université  Côte  d’Azur,  Observatoire  de  la  Côte  d’Azur,  CNRS,Laboratoire Lagrange, Bd de l’Observatoire, CS 34229, 06304 NiceCedex 4, France}

\maketitle\abstracts{I provide a summary of the ESA space astrometry mission Gaia regarding its main objectives and current status following the 2nd data release (Gaia DR2) in April 2018. The Gaia achievements in astrometry are assessed with a historical perspective by comparing the DR2 content to sky surveys or parallax searches over the last two centuries. One shows that Gaia sounds more like a big leap into a new world than an incremental progress in this field.}

\section{The Gaia mission}

\subsection{Context}
A major change in the precise measuring of position and displacement of celestial objects took place with the access to space. It is not exaggerated to refer to this period, starting with the selection in 1980 by ESA of the first space astrometric mission  Hipparcos, as a new golden age for this discipline which rests upon a long history of at more than two millennia, at least the Mediterranean world. Although Milky Way stars are the main celestial sources concerned by modern astrometry,  Gaia has more diverse targets with the positions and motions of $100,000$s of solar system bodies, stars within nearby external galaxies and the most remote quasars at the outskirts of the Universe. 

The extended review of Hipparcos results by Perryman  \cite{2012aaa..book.....P} amply demonstrates that the mission products influenced many areas of astronomy, in particular the structure and evolution of stars, the kinematics of stars and stellar groups. Even with its limited sample in number and variety of stars and observed volume, Hipparcos also made significant advances in our knowledge of the structure and dynamics of the Milky Way.

 Hipparcos was a resounding and acclaimed international success allowing the Europeans to quickly submit several more ambitious proposals for space astrometry, at the same time as others were also proposed to NASA or to the Japanese space agency \cite{1996A&AS..116..579L}. Only one of these proposals survived the various examinations by selection committees and Gaia was eventually selected as an ESA cornerstone mission in April 2000 for a launch around 2011. 
    
The basic observing concept is directly drawn from Hipparcos, but with a much larger telescope (actually two telescopes), a mosaic of 106 CCD detectors replacing the outdated photoelectric detector of Hipparcos. Two other instruments were added to carry out spectrophotometry and spectroscopic measurements, the latter to measure the velocity along the line of sight. While Hipparcos catalogue was limited to $100,000$ pre-defined stars brighter than 13.2 mag, Gaia was designed to realise a sensitivity-limited survey to $20$ mag. Hipparcos could only take a star at a time while Gaia is able to record simultaneously several $10,000$s  images mapped on its focal plane. About $1.5$ billion stars, amounting to $\approx 1$ percent of the
 Milky Way stellar content, are repeatedly observed during the nominal 5-year mission, leading to a final astrometric accuracy of $25$ {\muas} at $G = 15$ mag. (1 {\muas} = 0.001 mas = $10^{-6}$ arcsec). 
 
  \begin{table}[h]
\centering
        \caption{Overall content of the Gaia DR2. \label{table:summary}}\vspace{8pt}
   \begin{tabular}{lr}   
  \toprule        
  Data product or source type                                     & Number of sources               \\
  \midrule
  Total (excluding Solar system)                                  & {\ensuremath{1\,692\,919\,135}} \\[2pt]
  Five-parameter astrometry (position, parallax, proper motion)   & {\ensuremath{1\,331\,909\,727}} \\
  Two-parameter astrometry (position only)                        & {\ensuremath{361\,009\,408}}    \\
  ICRF3 prototype sources (link to radio reference frame)         & {\ensuremath{2\,820}}           \\
  Gaia-CRF2 extra-galactic sources (optical reference frame)     & {\ensuremath{556\,869}}         \\
  $G$-band (330--1050~nm)                                         & {\ensuremath{1\,692\,919\,135}} \\
  \ensuremath{G_\mathrm{BP}}-band (330--680~nm)                    & {\ensuremath{1\,381\,964\,755}} \\
  \ensuremath{G_\mathrm{RP}}-band (630--1050~nm)                   & {\ensuremath{1\,383\,551\,713}} \\[2pt]
    Median radial velocity over 22 months                           & {\ensuremath{7\,224\,631}}      \\
  Classified as variable                                          & {\ensuremath{550\,737}}         \\
  Variable type estimated                                         & {\ensuremath{363\,969}}         \\
  Detailed characterisation of light curve                        & {\ensuremath{390\,529}}         \\[2pt]
  Effective temperature {\ensuremath{T_\mathrm{eff}}}              & {\ensuremath{161\,497\,595}}    \\
  Extinction {\ensuremath{A_G}}                                   & {\ensuremath{87\,733\,672}}     \\
  Colour excess {\ensuremath{E(G_\mathrm{BP}-G_\mathrm{RP})}}       & {\ensuremath{87\,733\,672}}     \\
  Radius                                                          & {\ensuremath{76\,956\,778}}     \\
  Luminosity                                                      & {\ensuremath{76\,956\,778}}     \\[2pt]
  Solar system object epoch astrometry and photonetry             & {\ensuremath{14\,099}}          \\
  \bottomrule
  \end{tabular}          
 \end{table}

 \subsection{The mission main features}
 Gaia's main scientific goal is to clarify the origin and history of our Galaxy, from a quantitative census of the stellar populations and extremely accurate astrometric measurements to derive proper motions and parallaxes. See \cite{2001A&A...369..339P} for the proposal and \cite{2016A&A...595A...1G} for a presentation of the actual mission, the spacecraft, the operations and the data acquisition strategy. The principle of the scanning satellite relies on a slowly spinning spacecraft to measure the crossing times of stellar images transiting on the focal plane. As for Hipparcos, there are two fields of view combined onto a single focal plane where astrometric measurements are done. The time relates at once the one-dimensional star position to the instrumental axes. The relation to the celestial frame is obtained with the satellite attitude, which is solved simultaneously with the star positions in a global solution as described technically by Lindegren et al. \cite{2012A&A...538A..78L}. 
 
 \subsection{The mission and early results}
The Gaia satellite was launched on 19 December 2013 and the science data collection started after the in-flight qualification on 25 July 2014. A first batch of results was released on 15 September 2016 with only 14 months of data processed. This release comprised primarily a position catalogue (only two position parameters per source) for  1.14 billion stars, the largest such collection ever. A smaller catalogue combining Gaia and Hipparcos included parallaxes and proper motions for $\approx 2,000,000 $ stars with a sub-mas accuracy \cite{2016A&A...595A...2G}. The release contained also variable stars and a set of $2,200$ quasars common to Gaia and the radio ICRF   used to align the Gaia and the radio frame. Therefore the Gaia reference frame and ICRF are nominally identical.
 
However this first release (Gaia DR1) based on just above one year of data was far from touching the core objective of the mission, since the number of parallaxes available was very small, even though it was already about 20 times larger than with Hipparcos. The second release (Gaia DR2) came out on 25 April 2018. Gaia DR2 data is based on data collected between 25 July 2014 and 23 May 2016, spanning a period of 22 months of data collection, enough to disentangle the parallactic motion from the proper motion for most of the stars. The reference epoch for Gaia DR2 is J2015.5 and positions and proper motions are referred to the ICRS, to which the optical reference frame defined by Gaia DR2 is aligned. The main numbers describing the content are given in \ref{table:summary} and everyone would agree that they are impressive, even for those not well aware of how many of these stellar parameters were known before. Actually we will see in \ref{sect:significance} that the gap with existing data is really wide. An overall description of the content of this survey is given in a general paper from the Gaia collaboration \cite{2018A&A...616A...1G} from which I took the two main tables \ref{table:summary}-\ref{table:accuracy}.

 \begin{table}[h]
\centering
        \caption{Overall statistics of performance of the Gaia DR2. \label{table:accuracy}}\vspace{8pt}      
  \begin{tabular}{lr}   
  \toprule         
  Data product or source type                                               & Typical uncertainty\\
  \midrule 
  Five-parameter astrometry (position \& parallax)                          & $0.02$--$0.04$~mas at $G<15$\\
                                                                            & $0.1$~mas at $G=17$\\
                                                                            & $2$~mas at $G=21$\\ 
  Five-parameter astrometry (proper motion)                                 & $0.07$~mas~yr$^{-1}$ at $G<15$\\
                                                                            & $0.2$~~mas~yr$^{-1}$ at $G=17$\\
                                                                            & $3$~mas~yr$^{-1}$\ at $G=21$\\[4pt]
  Systematic astrometric errors (sky averaged)                              & $<0.1$~mas\\
  Gaia-CRF2 alignment with ICRF                                             & $0.02$~mas at $G=19$\\
  Gaia-CRF2 rotation with respect to ICRF                                   & $<0.02$~mas~yr$^{-1}$ at $G=19$\\[4pt]
 Mean $G$-band photometry                                                  & $0.3$~mmag at $G<13$\\
                                                                            & $2$~mmag at $G=17$\\
                                                                            & $10$~mmag at $G=20$\\
  Mean \ensuremath{G_\mathrm{BP}}- and \ensuremath{G_\mathrm{RP}}-band photometry   & $2$~mmag at $G<13$\\
                                                                            & $10$~mmag at $G=17$\\
                                                                            & $200$~mmag at $G=20$\\[4pt]

  Median radial velocity over 22 months                                     & $0.3$~km~s$^{-1}$ at $G_{\rm RVS} < 8$\\
                                                                            & $1.8$~km~s$^{-1}$ at $G_{\rm RVS} = 11.75$\\
  Systematic radial velocity errors                                         & $<0.1$~km~s$^{-1}$ at $G_{\rm RVS} < 9$\\
                                                                            & $0.5$~km~s$^{-1}$ at $G_{\rm RVS} = 11.75$\\[4pt]
  Effective temperature {\ensuremath{T_\mathrm{eff}}}                       & 324~K\\
  Extinction {\ensuremath{A_G}}                                             & 0.46~mag\\
  Radius                                                                    & 10\%\\
  Luminosity                                                                & 15\%\\[4pt]
  Solar system object epoch astrometry                                      & 1~mas (in scan direction)\\
  \bottomrule 
  \end{tabular}
\end{table}

 \section{The DR2 place in history}\label{sect:significance}
 Even at the level of the Gaia DR2, the Gaia products are just amazing in terms of volume and quality.  Although the numbers look impressive, but arguably this is true for many space missions, it is not straightforward to appreciate how much better or larger it stands compared to available astronomical data. I attempt to address this issue by showing that Gaia is producing much more than an incremental advance in the collection of basic astrophysical data, both in volume and accuracy, but that we are witnessing a true discontinuity of historical significance. For this purpose I compare the content of the Gaia survey to the astronomical knowledge at different times in the past. It is never easy to tell without ambiguity \textsl{how many parallaxes or proper motions} were known in 1850 or 1950. There is no clear  boundary of what should be considered as known and published values may be wrong or so poor as not providing a real knowledge. Therefore the figures on the set of tables below are often rounded and in details could be disputed, but not by large margin. At the end the trend is clear and does not suffer discussion.
 
 \subsection{The general trend}
My main point regarding the jump or discontinuity created by the Gaia catalogue is supported by the Table~\ref{table:history}. It gives for the most relevant parameters covering astrometry, spectroscopy and photometry the size of the available material at three different epochs, namely 1900, 1980, 2018. Although the number of stars for which one knows the parallaxes or the radial velocity say nothing about how well this is known, the large difference in the figures speaks for itself, and usually the increase in size goes with an increase in accuracy. \textsl{Sky survey} refers to the number of stars that have been catalogued with a position good enough to identify the star at any further epoch, a census of the sky in short.

\begin{table}[h]
\setlength{\tabcolsep}{5mm}
\centering
        \caption{General evolution in the size of general and specialised stellar catalogues. \label{table:history}}\vspace{8pt}
 \begin{tabular}{lccc}
 \toprule
Parameter  &\cent{1900}  &\cent{1980} & \cent{2018} \\
\midrule 
sky survey          &  800\,000  &  20 M$\,^a$   & 1.7 B \\[1pt]
fundamental sources &   350      & 1\,500$\,^b$    & 550\,000\\[1pt]
stellar parallaxes  &   100      & 10\,000  & 1.3 B \\[1pt]
proper motions      &   1000     & 250\,000$\,^c$ & 1.3 B \\[1pt]
radial velocities   &   50       & 30\,000  &  7.2 M \\[1pt]
variable stars      &   300      & 15\,000  &  500\,000 \\[1pt]
stellar temperature &   10\,000  & 300\,000 &  150 M\\[1pt]
\bottomrule
\footnotesize{$^a$ GSC-I,$^b$ FK5, $^c$ SAO }\\
\end{tabular}
\end{table}

\subsection{The Survey}
In comparing size of surveys at different epochs, I looked at the largest cataloguing of the whole sky, without caring about the positional accuracy. The object of these censuses was just to  list the place of the  sources on the sky up to a certain magnitude, with a positional accuracy good enough to recover the star and tell later between a known star and a moving body or a new star (becoming bright enough to be seen, but not really new and often referred to as a \textsl{guest star}). At different epochs efforts were made to get a completeness, at least from one place, of all the stars brighter than a given limit and provide also an approximate magnitude to help in the identification. The numbers of stars growing exponentially with the limiting magnitude,  these systematic surveys were quickly limited by the observational burden and the data handling, even when photographic plates became available. The compilation is given in Table \ref{table:stat_surveys} and provides an idea of how many stars appeared in the largest surveys in the past. By the end of XIXth century, all the stars to $V\approx 10$ were catalogued. The progresses were truly slow and most of the advances since 1950 were made possible with the advent of large computers with massive data storage. Gaia survey is today the largest, but smaller areas of sky have been investigated by optical or near IR telescopes at fainter magnitude than Gaia limit. In the coming years the LSST will outnumber Gaia by one order of magnitude. For Gaia  this was not a science goal to make the biggest survey in the sense it was done 150 years ago. However the last column of the table show where the strength of Gaia really lies in terms of positional accuracy.

\begin{table}[h]
\setlength{\tabcolsep}{5mm}
\centering
        \caption{Historical sky surveys. \label{table:stat_surveys}} \vspace{8pt} 
\begin{tabular}{lcrcr}
\toprule
\multicolumn{1}{c}{Name} & \multicolumn{1}{c}{Epoch}&\multicolumn{1}{c}{Num. stars} &  mag $< $ & \multicolumn{1}{c}{Accuracy} \\
\midrule
Ptolemy   &   II$^e$ siècle  & 1030       &  6       &   0.3 \degr  \\[1pt]
Flamsteed &   1700           & 2934       &  7       &   20 \arcsec \\[1pt]
Lacaille  &   1750           & 10\,000    &  7       &   5   \arcsec\\[1pt]
Lalande   &   1800           &  48\,000   &  9       &   3  \arcsec \\[3pt]
BD        &   1860           & 450\,000   &  9.5     &   20  \arcsec \\[1pt]
CD        &   1900           & 600\,000   &  10      &   20  \arcsec \\[1pt]
CPD       &   1900           & 450\,000   &  11      &    5   \arcsec \\[1pt]
Carte du Ciel & 1950         & 2 M        &  12      &   0.5  \arcsec \\[1pt]
Tycho 2   &   2000           & 2.5 M        &  11.5    &   50 mas    \\[3pt]
2MASS     &   2000           & 300 M      &  15 (IR) &   500   mas \\[1pt]
USNO B1   &   2000           & 1000 M     &  21      &   200  mas \\[1pt]
\textbf{Gaia }     &   2015           & 1700 M     &  21      &   0.2   mas \\[1pt]
LSST      &   2030 ?         &  15 B      &  24      &    5 mas  \\ 
\bottomrule
\end{tabular}
\end{table}

\subsection{The reference system}
For the general public a catalogue of stars is before all a list of sources one can locate in the sky thanks to the two angular coordinates given in each line. This is more or less the goal of the sky census just mentioned. But scientifically more important is the use of an accurate position catalogue to materialise a reference frame in space. As said by W. Fricke\cite{1985CeMec..36..207F}, a recognised expert in this field, \textsl{the main purpose of a Fundamental Catalogue in astronomy consists in providing a coordinate system for describing the motion of the planetary systems   $(\cdots$)  and for the determination of the proper motions of stars}. In short one wishes to have consistent coordinates for a small set of point-sources on the sky so that this can be used to fix the three orthogonal directions $[x, y, z]$ of an inertial system, although the language should be slightly altered in the framework of General Relativity. Producing a fundamental catalogue requires absolute observations not depending on earlier positions of the same stars or of other stars used to derive relative positions. For the general principles and construction see the book by H.G. Walter and O.J. Sovers \cite{2000afce.conf.....W}. Building a fundamental catalogue is a very demanding task, extending over years and even decades, and a handful have ever existed containing few 100-1000s sources, stars until Hipparcos and quasars more recently.

Table~\ref{table:stat_refcat} provides an almost exhaustive list of fundamental catalogues since the concept has emerged in the mid-XVIIIth century. Comparing the 3rd column in this table and in Table \ref{table:stat_surveys} is enough to grasp the  wide difference between the two kinds of catalogues, two different worlds indeed. Remarkably there is only one catalogue common  to these two tables: Gaia, but not for the same sources. The Gaia CRF (Celestial Reference Frame) is built on a small subset of sources, namely the $550\,000$ QSOs meeting the overall principles of the ICRS (International Celestial Reference System), as explained in Mignard et al. \cite{2018A&A...616A..14G}. Gaia accuracy is on a par with the most recent version of the ICRF (the IAU approved fundamental reference frame) in the radio wavebands and  resulting from VLBI astro-geodetic observations. However the density is 100 times higher and it is directly accessible in the optical domain. Both realisations are needed and used in different contexts in astronomy and geodesy.
\begin{table}[h] 
\setlength{\tabcolsep}{3mm}
\centering
        \caption{Precision of the historical fundamental catalogues. \label{table:stat_refcat}} \vspace{8pt} 
\begin{tabular}{lcrcr}
\toprule
\multicolumn{1}{c}{Name} & \multicolumn{1}{c}{Epoch}&\multicolumn{1}{c}{Sources} &  mag $< $ & \multicolumn{1}{c}{Accuracy} \\
\midrule
Lacaille       & 1760       & 397        &  7      &   10   \arcsec \\[1pt]
Maskelyne      & 1774       & 36         &  5      &    5  \arcsec \\[1pt]
Piazzi         & 1818       & 220        &  6      &    2  \arcsec \\[1pt]
Bessel         & 1830       & 36         &  5      &    1  \arcsec \\[1pt]
Argelander     & 1869       & 160        &  6      &    1  \arcsec \\[3pt]
Auwers         & 1879       & 539        &  6      &   0.5  \arcsec \\[1pt]
FK4            & 1963       & 1535       &  7.5    &   0.2 \arcsec \\[1pt]
FK5            & 1988       & 1535       &  7.5    &   40 mas      \\[3pt]
Hipparcos$^a$  & 1996      & 100\,000    &  11.5   & 1 mas    \\[3pt]
ICRF1  (radio) & 1998       &  620       &  -      &    2 mas    \\[1pt]
ICRF2  (radio) & 2009       &  3400      &  -      &    0.6 mas  \\[1pt]
ICRF3  (radio) & 2018       &  4500      &  -      &    0.2 mas  \\[3pt]
\textbf{Gaia } QSOs & 2018  &  550\,000  &  21  & 0.4 mas  \\[1pt]
\textbf{Gaia } QSOs $G< 18$ & 2018  &  27\,000  &  18  & 0.12 mas  \\
\bottomrule
\footnotesize{$^a$  quasi-fundamental }\\
\end{tabular}
\end{table}

\subsection{The parallaxes}
Photometry, radial velocities, stellar parameters can in principle be obtained with ground based instruments, provided the necessary means and human resources are made available over several years to support this routine work. This is \textsl{easier} and more productive from space, but space is not technically required. This is not true for absolute parallaxes which need the finest astrometry not achievable through the atmosphere. The parallaxes of stars, or equivalently their distances, were searched as soon as the Heliocentric system became accepted by astronomers. Our displacement around the Sun should translate into an annual reflex motion of the stars whose extent is in direct relation to their distances. So the underlying geometric principles needed to ascertain the distances  are extremely simple and were well understood well before any realistic attempt to detect the tiny parallactic effect was feasible. The first success cam with F. Bessel in 1838 and further progress were extremely slow. Parallaxes are small in regard of the achievable astrometric accuracy and their astrophysical value is low if not better than $10\%$ in fractional accuracy.  Even in 1980, just before Hipparcos, many of the 8000 stars with published parallaxes did not reach this goal. In addition they were relative parallaxes and depended on additional calibration/assumptions for the reference stars used to record the parallactic displacement. 

\begin{table}[h]
\setlength{\tabcolsep}{5mm}
\centering
        \caption{Number of  published stellar parallaxes.\label{table:stat_varpi}} \vspace{8pt}
   \begin{tabular}{lrl}
   \toprule
   year  &\cent{number of stars}  &\cent{comment}   \\[2pt]
        \midrule          
 1840  &3 & 61 Cygni, Vega, $\alpha$ Cent\\[2 pt]
 1850    & 20 &  C.A.F. Peters \\[2 pt]
 1890    & 45 & Ch. André (astron. stell.) \\[2 pt]  
 1910    & 300& incl. 52 with photography \\[6 pt]
 1925    & 2000 & photographic plates\\[2 pt]
 1965    & 6000 & Yale Catalogue \\[2 pt]
 1990    & 8000 & just before Hipparcos\\[6 pt]
 1996    & 120\,000 & Hipparcos \\[2 pt]
 \textbf{2018}    & 1\,300 M   & Gaia \\
\textbf{2025}    & 1\,800 M  & Gaia \\
 \bottomrule
\end{tabular}
\end{table}

Hipparcos astrometry was a truly new start for parallax survey. The total number of trigonometric parallaxes rose at once to more than $100,000$, with nearly $50,000$ better than 20\% in fractional errors ($\sigma_\varpi/\varpi < 20\%$)  and $20,000$ at the $10\%$ level. The new technique, allowing to get virtually absolute parallaxes, proved efficient and multiplied in only 3 years the number of measured parallaxes by a factor 15. This success paved the way for Gaia and the numbers given in Table \ref{table:stat_varpi} demonstrate without further comments the power of Gaia compared to the Ground-based methods, painstaking and with low yield. This is a true revolution in this field, something that any astronomer of the 1980s could not simply dream of. With Gaia DR2 one has parallax estimates for $1.3$ billion stars, but more important above 50 million  with a relative accuracy better than $10\%$. With the coming releases this figure will soar above the 100 million. The Gaia parallax survey is by far the most comprehensive ever done and has no match in terms of size and accuracy, with the exception, regarding the accuracy, of a handful of radio masers observed with the VLBI technique. For the Galactic stars this is the crowning of nearly two centuries of parallax quest starting with F.W. Bessel in 1838. Until many years in the future there will be no such undertaking to get trigonometric parallaxes directly from astrometric observations and the Gaia survey is now actively being used to reconstruct the whole distance scale beyond the Galaxy, based on secondary indicators.

\subsection{The proper motions}
The same historical comparison is done for the proper motions of stars with results given in Table~\ref{table:stat_pm}. Detecting the displacement of stars is in principle not too difficult, since it required basically only patience and archives. Patience, since the changes in the relative positions of stars is very slow at the scale of human life, and then one must wait decades or centuries before it becomes manifest. But for that one needs also to keep track of where the stars were in the past, otherwise no comparison is feasible. It was not until the early years of the XVIIIth century that hints of a possible break in the doctrine of the \textsl{fixed stars} came to the fore and no indisputable quantitative measurement was available before J. Cassini confirmed beyond doubt that Arcturus has changed position over 40 years. Table~\ref{table:stat_pm} is rather similar to that for the parallaxes and just illustrates the sluggish progress of astrometry until space astrometry became a reality. Accuracy is a key factor, since the computed positions degrade in proportion of the uncertainty of the proper motions. The impact of Gaia is obvious with an incredible increase in quantity and quality. And the icing on the cake with Gaia, angular displacement combined with parallaxes gives the true velocity in {\kms}. And for the bright end, Gaia spectrometer delivers also the third component on the line of sight.

\begin{table}[t]
\setlength{\tabcolsep}{5mm}
\centering
        \caption{Number of  known proper motions at different epochs.\label{table:stat_pm}} \vspace{8pt}
 \begin{tabular}{lrl}
 \toprule
 Year  &\cent{num. stars}  & \cent{comment}\\[2pt]
\midrule 
 1738    & 1         & J. Cassini  \\[2 pt]       
 1760    &15        &  T. Mayer \\[2 pt]
 1790    &36        &  N. Maskelyne \\[2 pt]
 1835    & 390      &  Argelander, $\mu >  100  \text{ mas/a}$ \\[2 pt]
 1856    & 3200     &  Mädler, (Bradley's stars) \\[6 pt]
 1939    & 33\,000  &   Gal. Cat., $ V < 7.5$ $\sigma \approx 10 \text{ mas/a}$ \\[2 pt]
 1966    & 260\,000  & SAO $\sigma \approx 15 \text{ mas/a}$ \\[6 pt]
 1990    & 400\,000  & PPM $\sigma \approx 4 \text{ mas/a}$ \\[2 pt]
 1996    & 120\,000  & Hipparcos $\sigma \approx 1  \text{ mas/a}$  \\[2 pt]
 2002    & 2.5 M & Tycho-2 $\sigma \approx 2.5  \text{ mas/a}$  \\[6 pt]
 \textbf{2018} & 1.3 B   & Gaia $\sigma \approx 0.3  \text{ mas/a} \text{ @ } G =18$ \\
 \textbf{2025} & 1.7 B   & Gaia $\sigma \approx 0.03  \text{ mas/a} \text{ @ } G =18$ \\
\bottomrule
\end{tabular}
\end{table}

\section*{Acknowledgments}
The section summarising the Gaia mission takes heavily on the Gaia on-line documentation and summary papers prepared by the DPAC collaboration to accompany the data releases. The members of the DPAC are collectively thanked for providing this material and for their dedication to the success of Gaia.

\section*{References}
\bibliography{FM_DS_bibfile,BiblioICRF}

\end{document}